\documentclass[12pt,twocolumn]{iopart}

\usepackage{graphicx}
\usepackage[font=footnotesize,format=plain,labelfont=bf]{caption}

\begin{document}

\title{Quantum process tomography of single photon quantum channels with controllable decoherence}

\author{A Shaham and H S Eisenberg}

\address{Racah Institute of Physics, Hebrew University of
Jerusalem, Jerusalem 91904, Israel}
 \ead{assaf.shaham@mail.huji.ac.il}

\begin{abstract}
We characterized unital quantum channels of single photon
polarization qubits. The channels are composed of two birefringent
crystals and wave-plates, where their decoherence properties are
controlled. An experimental comparison between two different
depolarizing configurations was performed using a quantum process
tomography procedure. The results are with a high fidelity to the
theoretical predictions.
\end{abstract}

\submitto{\PS}

\maketitle

In recent years, within the growing interest in quantum
communication, many schemes have used the polarization of a single
photon to encode a two level quantum system - a qubit
\cite{Nielsen}. Interactions between such a photon and other
surrounding systems may result in a decoherence process where noise
is added to the quantum information that is stored in the photon
polarization. The possibility to control and characterize the
properties of this noise enables the study of its effects on various
quantum protocols. In particular, it will be possible to test the
performance of quantum error correction schemes and the efficiency
of quantum key distribution protocols \cite{BB84,Bruss}. The
construction of a quantum channel that can induce different types of
noises is important also for the study of a quantum-classical
transition that a system may experience \cite{Almeida,Xu}.

Implementing a channel with controlled depolarization properties is
a challenging task, since light interacts very weakly with its
environment. Such a channel can be realized with the aid of
birefringent crystals which entangle the polarization degree of
freedom with the photon's temporal degrees of freedom \cite{Kwiat}
or with its spatial degrees of freedom \cite{Nambu,Puentes}. In this
work, we studied a depolarizing channel which couples between
polarization and time of light of short coherence times
\cite{Shaham}. It is composed of two birefringent crystals and wave
plates. The coupling depends on the relative angle between the
symmetry axes of the crystals, the wave-plate orientations and the
initial polarization of the input state. Depolarization occurs
because photon detection is insensitive to the temporal degrees of
freedom.

The polarization state of a single-photon qubit can be described
either by a density matrix operator $\hat{\rho}$ or, equivalently,
by a point in the Poincar\'{e} sphere. The Cartesian coordinates of
this point are the Stokes parameters $\overline{S}=\{S_1,S_2,S_3\}$.
Here we use the convention that $S_1$ represents the linear
horizontal and vertical polarizations ($|h\rangle$,$|v\rangle$),
$S_2$ represents the linear polarizations plus and minus $45^\circ$
($|p\rangle=(|h\rangle+|v\rangle)/\sqrt{2}$,
$|m\rangle=(-|h\rangle+|v\rangle)/\sqrt{2}$), and $S_3$ represents
the circular polarizations
($|r\rangle=(|h\rangle+i|v\rangle)/\sqrt{2}$ and
$|l\rangle=(i|h\rangle+|v\rangle)/\sqrt{2}$). The length of the
Stokes vector $D=\sqrt{S^2_1+S^2_2+S^2_3}$ represents the state's
degree of polarization. For polarized states, $D=1$, while for
partially polarized states, $D<1$. Characterization of the
polarization state may be performed by several projection
measurements with the Quantum State Tomography (QST) procedure
\cite{Altepeter_QST}.

Consider an arbitrary quantum state $\hat{\rho}$ that serves as an
input state to a quantum channel. The final state $\hat{\rho}'$ is
derived from the mapping $\hat{\rho}'=\mathcal{E}({\hat{\rho}})$
where $\mathcal{E}$ is the operation of the channel. The mapping
$\mathcal{E}$ can be uniquely described by the elements of the
positive and Hermitian matrix $\chi$:
\begin{equation}\label{process matrix}
\mathcal{E}(\hat{\rho})=\sum_{m,n}\chi_{mn}\hat{E}_{m}\hat{\rho}\hat{E}_{n}^\dag,
\end{equation}
where $\hat{E}_m$ are basis elements that span the space of
$\hat{\rho}$. The elements of the $\chi$ matrix can be
experimentally determined by a Quantum Process Tomography (QPT)
procedure \cite{Chuang_QPT}. A QPT procedure for a single-qubit
process requires four QST measurements of different input states. We
note that these four states should not lay on the same plane in the
Poincar\'{e} sphere representation. Assuming the process has some
special symmetries may reduce the number of QST measurements that
are required for the reconstruction of the $\chi$ matrix. A complete
positive and trace preserving process is a unital process if
$\mathcal{E}(\hat{I})=\hat{I}$. A unitality assumption for a single
qubit process may reduce the number of QST measurements that are
needed for the process chracterization to 3.

\begin{figure}[tbp]
\centering{}
\includegraphics[angle=0,width=86mm]{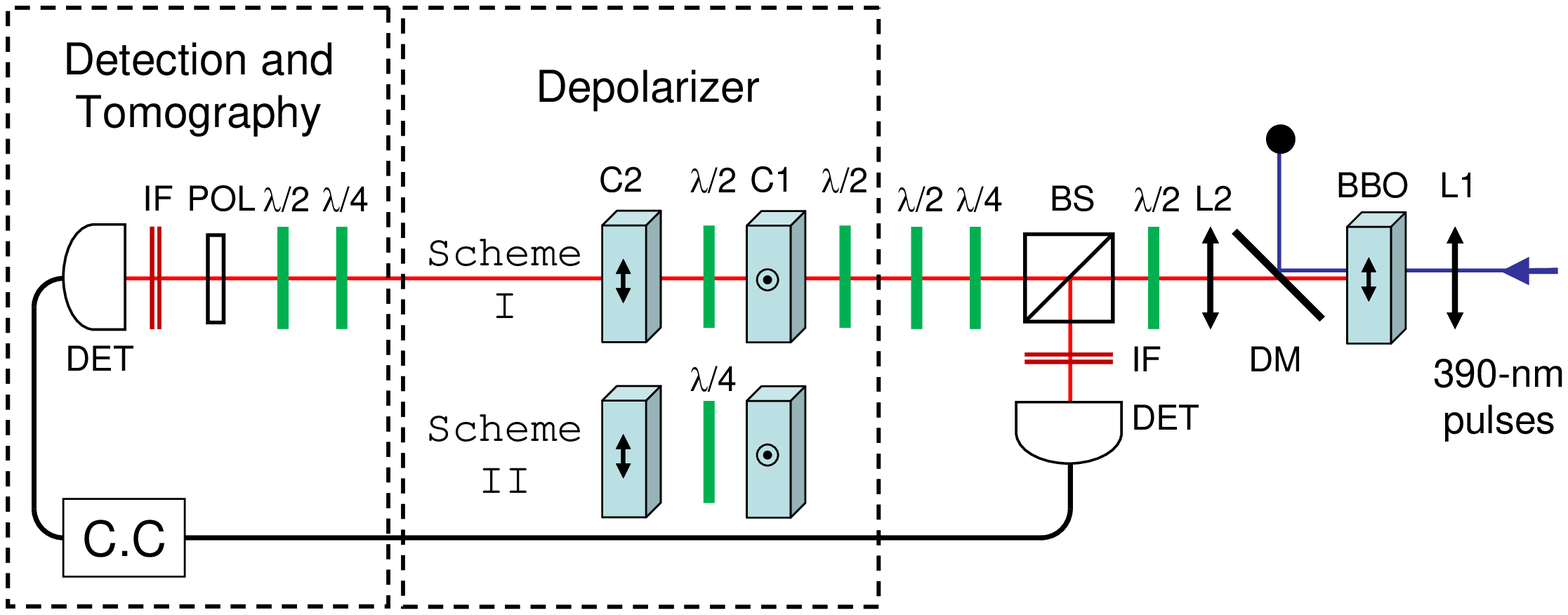}
\caption{\label{setup} (Color online) The experimental setup and the
depolarizing schemes. Laser pulses of 390 nm were focused by a lens
(L1) into a BBO crystal and reflected by a dichroic mirror (DM).
Down-converted signal was collimated by another lens (L2). One
photon of the down-converted pair was probabilistically split by a
beam splitter (BS) and detected by a single-photon detector (DET).
The initial polarization state of the second photon was determined
using half- and qurter-wave plates ($\lambda/2, \lambda/4$). The
signal was directed to one of the two depolarizing schemes as shown.
QST measurements were carried out using wave plates, a polarizer
(POL) and another detector. The photons were temporally filtered
using a 5 nm bandpass interference filters (IF). Spatial filtering
was achieved with the coupling of the photons to single-mode fibres
connected to the detectors. }

\end{figure}

The aim of this work is to study theoretically and experimentally
the processes of two depolarizing channel schemes. The first channel
(Scheme I, see figure \ref{setup}) is composed of two equal
birefringent crystals and two half-wave plates. The crystals are
fixed perpendicularly with respect to each other, and the rotatable
wave plates are placed before and after the first crystal. The
second channel (Scheme II, see figure \ref{setup}) is composed also
of two equal perpendicularly fixed birefringent crystals, but here a
quarter-wave plate is placed in between them. For both depolarizing
schemes the symmetry axes of the crystals define the linear
$|h\rangle$ and $|v\rangle$ polarizations, and the zero angle of the
wave plates is determined with respect to the horizontal
polarization. The coupling between polarization and temporal modes
of such a channel can be described as follows: every crystal induces
a temporal delay $t$ between the horizontally and vertically
components of an input polarized wave packet. This temporal walk-off
$t$ depends on the crystal length L, its refractive index difference
$\Delta n$ and the speed of light $c$ such that $t= L\frac{\Delta
n}{c}$. In order to achieve a complete separation (i.e. loss of
coherence between horizontal and vertical wave packets), we require
that $t>t_c$, where $t_c$ is the coherence time of the initial wave
packets. We see that if there is no polarization rotation between
the two perpendicular crystals of our schemes (i.e. a wave plate
that is placed in between them is set to a zero angle), the second
crystal compensates for the time delay that was caused by the first
one and no depolarization occurs. If there is a polarization
rotation between the two crystals, two or three different temporal
modes will be occupied after the second crystal. Tracing out these
temporal degrees of freedom while measuring the polarization state
will result in depolarization. It is easy to calculate the processes
that these channels induce for any angle of the wave plates because
in both schemes the polarization is coupled to a maximum of three
discrete temporal modes (for details see \cite{Shaham}). We
emphasize that the wave plates act on any of these modes separately;
thus they do not couple between the different temporal modes.
Theoretical calculations of the processes for both schemes have
shown that the processes are unital regardless of the orientation of
any wave plate in the channel configurations.

Our experimental setup is shown in figure \ref{setup}: a Ti:Sapphire
pulsed laser with a 76 MHz repetition rate was frequency doubled in
order to generate pulses with a wavelength of 390 nm. These pulses
were focused into and collinearly down-converted in a 1 mm thick
type-I BBO crystal. The 780 nm down-converted signal was filtered by
a dichroic mirror (DM) and collimated with a lens (L2). One photon
of the pair was probabilistically split by a beam splitter (BS) and
was sent to a detector. The second photon was directed to the
depolarizing channel. For the two depolarizing schemes, the
birefringent phase of the depolarizing crystals was tuned so that
the entire depolarizing channel will not affect the state when the
wave plate angles are set to zero. Characterizing the polarization
state of the depolarized photons was performed by wave plates and a
polarizer (POL). Photons were filtered by 5 nm bandpass filters
(IF), corresponding to a coherence time of $t_{c}\simeq180 \textrm{
fs}$, and then coupled into single-mode fibres, leading to
single-photon detectors (DET). The detection of the second photon
was conditioned by the detection of the first one in order to ensure
that the depolarized signal is truly a quantum state.

The probabilistic nature of a quantum polarization state, together
with systematic errors during the projection measurement procedure,
may result in an illegal linear reconstruction of the measured state
or the measured process. A Maximal Likelihood (ML) search that
restricts the parameters to physically allowed values can be used to
find the physical representation of $\hat{\rho}$ or $\chi$ that best
fits to the tomographic data. We reconstructed all of the state
density matrices using the ML protocol suggested by James {\it et
al} \cite{James_QST}. As for the $\chi$ matrices, although ML
searches for photonic QPT measurements were previously suggested and
demonstrated \cite{O'Brien_QPT,Jezek_QPT}, we used a simpler ML
search that gave sufficient results. According to the
Choi-Jamio{\l}kowski isomorphism, the linear mapping of a quantum
channel represents a legal quantum state of a larger dimension
\cite{Jamiolkowski}. Thus, we can assume that the linearly
reconstructed process matrix $\chi_{linear}$ represents a
four-dimensional (4D) quantum state. Additionally, we assume that
this state was reconstructed from a set of 16 probability
measurements $\{p_1,p_2,..,p_{16}\}$ obtained from 16 artificial
projection measurements on the states
$\{|\psi_1\rangle,...,|\psi_{16}\rangle\}$. Inverting the linear
reconstruction of a density matrix from the 16 probability
measurements, we calculated these artificial probability values from
$\chi_{linear}$. Then, we reconstructed a new $\chi$ matrix using
the artificial probability values with an ML quantum state search.
We emphasis that $\chi$ might not be the process with the best fit
to the data, but only a close physical fit to the originally
measured data.

As this ML search requires the translation of the artificial
probability measurements to a set of artificial counts, there is an
extra free multiplying parameter $\mathcal{N}$. For example, this
parameter can represent the measurement time. We chose $\mathcal{N}$
such that the artificial set of counts will have the same order of
magnitude as the actual qubit measured counts, which was about
25,000. Changing $\mathcal{N}$ had a minute effect on the $\chi$
matrix: multiplying or dividing $\mathcal{N}$ by $10^3$ resulted in
$\chi$ matrices whose fidelities with the original $\chi$ matrix
($F(\chi_1,\chi_2)=(\Tr{\sqrt{\sqrt{\chi_1}\chi_{2}\sqrt{\chi_1}}})^2$)
were higher than $99\%$. An error estimation of $2\%$ for the
process matrix fidelities and their eigenvalues resulted mainly from
this fact.

\begin{figure}[tbp]
\centering{}
\includegraphics[angle=0,width=60mm]{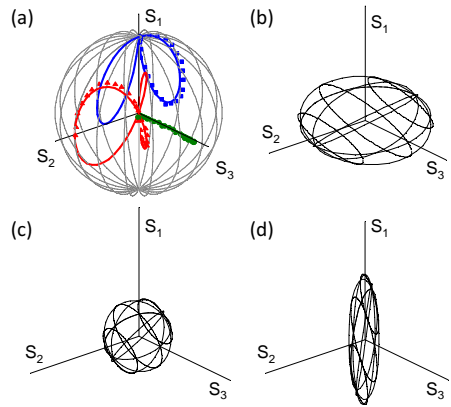}
\caption{\label{scheme_I_processes} (Color online) The first
depolarizing scheme: experimentally measured final states and QPT in
the Poincar\'{e} sphere representation. (a) Comparison of final
state measurements for input states of  $|h\rangle$ (blue squares),
$|p\rangle$ (red triangles), and $|r\rangle$ (green circles) in the
range $0^\circ<\theta<90^\circ$ and the theoretical model.
Theoretical curves are presented as solid lines in the range
$0^\circ<\theta<180^\circ$. (b-d) Mapping of the surface of the
Poincar\'{e} sphere to depolarized wire-mesh ellipsoids that were
obtained by experimental QPT for the crystal angle values of (b)
$\theta=35.3^\circ$, (c) $\theta=54.74^\circ$, and (d)
$\theta=67.5^\circ$.}
\end{figure}

Beginning with the first scheme, we rotated the two half-wave plates
in opposite directions by an angle of $\theta/2$. This rotation is
equivalent to the rotation of the first crystal by an angle of
$\theta$. We generated initial $|h\rangle$, $|p\rangle$ and
$|r\rangle$ polarization states, sent them through the depolarizer
and performed QST to the depolarized states for different $\theta$
angles in the range of $0<\theta<90^\circ$. The traces of the
measured states in the Poincar\'{e} sphere representation are shown
in figure \ref{scheme_I_processes}(a). Solid lines represent the
theoretical calculations for angles in the range
$0<\theta<180^\circ$. Figures \ref{scheme_I_processes}(b)-(d)
present the measured process of the channel mapping of the
Poincar\'{e} sphere surface when
$\theta=35.3^\circ=\tan^{-1}(\frac{1}{\sqrt{2}})$,
$\theta=54.7^\circ=\tan^{-1}(\sqrt{2})$, and $\theta=67.5^\circ$,
respectively. The initial states that were used for the QPT were
$|h\rangle$, $|p\rangle$ and $|r\rangle$, where the process of
$\theta=35.3^\circ$ was reconstruct using the unitality assumption,
and the other two processes were measured using the additional input
state of $|v\rangle$. The first process (figure
\ref{scheme_I_processes} (b)) maps the surface of the Poincar\'{e}
sphere to an ellipsoid with two primary radii of length
$\frac{2}{3}$ and the third one has a length of $\frac{1}{3}$. The
second process (figure \ref{scheme_I_processes} (c)) is an isotropic
depolarizing process that maps the polarized states to a sphere with
radius equal to $\frac{1}{3}$. The third process (figure
\ref{scheme_I_processes} (d)) corresponds to a mapping of the
Poincar\'{e} sphere surface to an ellipsoid with one primary radius
with a length of $\sim0.7$ and two primary radii with a length of
$\sim0.15$.

In order to evaluate the channel performance with respect to the
theoretical prediction, we compared between the measured $\chi$
matrices and the theoretically calculated $\chi_d$ matrices. A
comparison between the three measured maps that are presented in
figure \ref{scheme_I_processes}(b)-(d) and the theoretically
calculated maps resulted with fidelities $F$ higher than $97\pm2\%$.

\begin{figure}[tbp]
\centering{}
\includegraphics[angle=0,width=86mm]{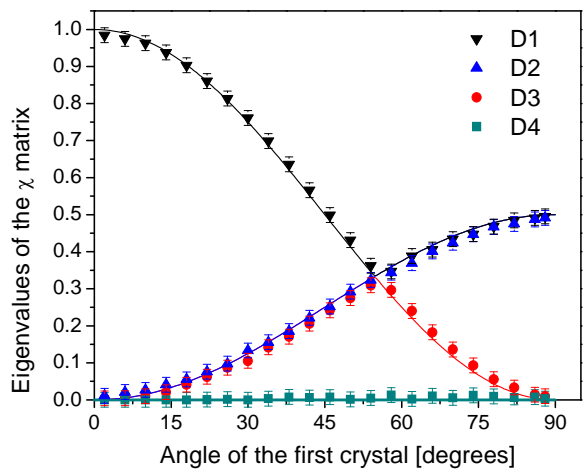}
\caption{\label{scheme_I_eigenvalues} (Color online) The
eigenvalues of the $\chi$ matrix as a function of the equivalent
first-crystal rotation angle $\theta$ for the first depolarizing
scheme.}
\end{figure}

A systematic study of the decoherence that the first scheme can
induce can be performed by exploring the eigenvalues of the 4D
$\chi$ matrix as a function of the angle $\theta$. Since they are
not affected by rotations and reflections of the input state, these
eigenvalues may serve as a measure of the pure decoherence types
that a quantum state may experience while passing through the
channel. Using the assumption of unitality, we have used the
transformed states that are presented in figure
\ref{scheme_I_processes}(a) to reconstruct the $\chi$ matrix for the
range of angles $0<\theta<90^\circ$. The measured eigenvalues of
these matrices are presented in figure \ref{scheme_I_eigenvalues}
along with their theoretical predictions. We see that only three
eigenvalues participate in the process, while the fourth eigenvalue
remains zero for every angle. Two eigenvalues out of these three are
always equal. When the process is isotropic ($\theta=54.7^\circ$)
the three eigenvalues intersect and are equal to $\frac{1}{3}$.

\begin{figure}[tbp]
\centering{}
\includegraphics[angle=0,width=86mm]{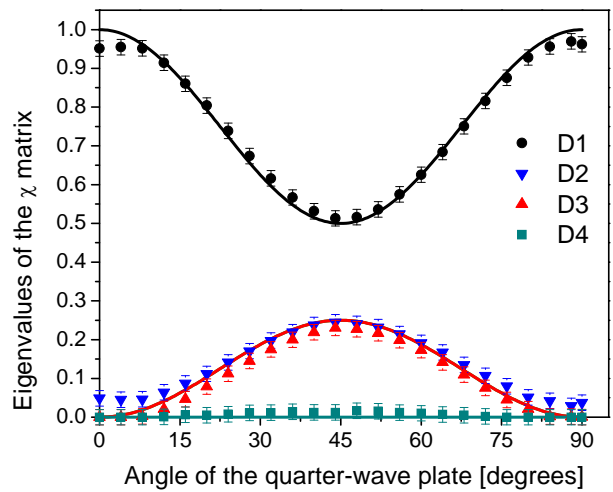}
\caption{\label{scheme_II_eigenvalues} (Color online) The
eigenvalues of the $\chi$ matrix as a function of the quarter-wave
plate angle $\phi$ for the second depolarizing scheme.}
\end{figure}

\begin{figure}[tbp]
\centering{}
\includegraphics[angle=0,width=85mm]{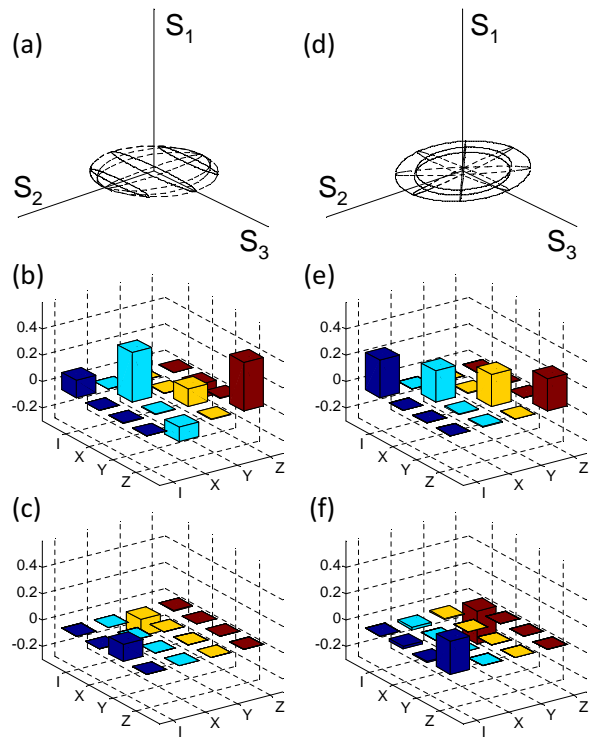}
\caption{\label{processes_45_deg} (Color online) A comparison
between the disk mapping processes of the first scheme and the
second scheme. (a-c) Measured mapping of the surface of the
Poincar\'{e} sphere, real and imaginary parts of the $\chi$ matrix
for the first scheme process of $\theta=45^{\circ}$. (d-f) The
corresponding results for the $\phi=45^{\circ}$ second scheme
process.}
\end{figure}

Moving to the second depolarizing scheme, we characterized the
channel process for different angles $\phi$ of the quarter-wave
plate in the range of $0<\phi<90^\circ$. The QPT was performed by
three QST measurements for outputs of the three initial mutually
unbiased states $\overline{S}_a=\{\sqrt{1/3},0,-\sqrt{2/3}\}$,
$\overline{S}_b=\{\sqrt{1/3},\sqrt{1/2},\sqrt{1/6}\}$, and
$\overline{S}_c=\{\sqrt{1/3},-\sqrt{1/2},\sqrt{1/6}\}$. It is worth
mentioning that these three states have the property that their
final output state after passing through the second scheme has the
same degree of polarization regardless of the angle $\phi$. Using
the assumption of unitality, we reconstruct the $\chi$ matrices of
the different processes. The eigenvalues of these matrices and their
theoretical predictions are presented in figure
\ref{scheme_II_eigenvalues}. As in the first scheme, only three
eigenvalues participate in the process, and two out of them are
equal for any orientation of the quarter-wave plate. Maximal
decoherence is obtained when $\phi=45^{\circ}$, where the
eigenvalues are $\frac{1}{2}$, $\frac{1}{4}$, $\frac{1}{4}$ and 0.
As can be seen in figure \ref{scheme_I_eigenvalues}, the same
eigenvalues of the $\chi$ matrix are obtained also with the first
scheme when $\theta=45^{\circ}$. A comparison between the
measurements of these two processes is shown in figure
\ref{processes_45_deg}: the mapping of the surface of the
Poincar\'{e} sphere according to scheme I process at
$\theta=45^{\circ}$ and the real and the imaginary parts of the
$\chi$ matrix are shown in figure \ref{processes_45_deg}(a)-(c),
respectively. The mapping and the real and imaginary part of the
$\chi$ matrix that are shown in figure \ref{processes_45_deg}(d)-(f)
correspond to the scheme II process at $\phi=45^{\circ}$. Both
processes map the Poincar\'{e} sphere into the shape of a disk, but
the rotations that accompany the processes are different. This is
reflected in the difference between the elements of the $\chi$
matrices and in the difference in the orientation of the wire mesh
of the mapping. The fidelities of these two measured processes to
the desired ones are $97\pm2{\%}$ and $98\pm2{\%}$ for the first
scheme process at $\theta=45^{\circ}$, and the second scheme process
at $\phi=45^{\circ}$, respectively.

In conclusion, we characterized the processes of two depolarizing
schemes composed of two perpendicularly fixed equal birefringent
crystals, and rotatable wave plates. All processes were analyzed
according to the eigenvalues of their $\chi$ matrix which serve as a
measure for the decoherence properties. The schemes realize
processes in which three eigenvalues of the $\chi$ matrix differ
from zero, and two out of them are equal. A comparison of scheme I
which contains a half-wave plate in between the crystals and scheme
II which contains a quarter-wave plate in between them reveals that
scheme I offers more possible decoherence types than Scheme II. Both
schemes exhibit high fidelities to the theoretical predictions.
Ongoing research on the decoherence of more composite quantum
systems is currently in progress using these depolarizing schemes.

The authors thank Jan Soubusta and Karel Lemr for a helpful
discussion, and the Israeli Science Foundation for supporting this
work under Grant No. 366/06.

\end{document}